\documentclass[preprint,12pt]{elsarticle}

\usepackage{amssymb}
\usepackage{amsmath}
\usepackage{graphicx}% Include figure files
\usepackage{dcolumn}% Align table columns on decimal point
\usepackage{bm}% bold math
\usepackage{braket}

 \usepackage[utf8]{inputenc}
\newcommand{\bea}{\begin{eqnarray}}
\newcommand{\eea}{\end{eqnarray}}
\newcommand{\simgt}{\hbox{ \raise3pt\hbox to 0pt{$>$}\raise-3pt\hbox{$\sim$} }}
\newcommand{\simlt}{\hbox{ \raise3pt\hbox to 0pt{$<$}\raise-3pt\hbox{$\sim$} }}

\newcommand{\be}{\begin{equation}}
\newcommand{\ee}{\end{equation}}

\newcommand{\LO}{\xi}
\newcommand{\NLO}{\xi^2}

\journal{Physics Letter B}

\begin{document}

\begin{frontmatter}

%% Title, authors and addresses

%% use the tnoteref command within \title for footnotes;
%% use the tnotetext command for theassociated footnote;
%% use the fnref command within \author or \address for footnotes;
%% use the fntext command for theassociated footnote;
%% use the corref command within \author for corresponding author footnotes;
%% use the cortext command for theassociated footnote;
%% use the ead command for the email address,
%% and the form \ead[url] for the home page:
%% \title{Title\tnoteref{label1}}
%% \tnotetext[label1]{}
%% \author{Name\corref{cor1}\fnref{label2}}
%% \ead{email address}
%% \ead[url]{home page}
%% \fntext[label2]{}
%% \cortext[cor1]{}
%% \address{Address\fnref{label3}}
%% \fntext[label3]{}

\title{ 
Probing Higgs self-coupling
of a classically scale invariant model\\
in $e^+e^- \to Zhh$:
Evaluation at physical point
}

%% use optional labels to link authors explicitly to addresses:
%% \author[label1,label2]{}
%% \address[label1]{}
%% \address[label2]{}

\author{Y.~Fujitani and Y.~Sumino}

\address{
Department of Physics, Tohoku University,
Sendai, 980--8578 Japan
}

\begin{abstract}
%% Text of abstract
A classically scale invariant extension of the standard model
predicts large anomalous Higgs self-interactions. 
We compute missing contributions in 
previous studies for probing the Higgs triple coupling
of a minimal model using the process $e^+e^- \to Zhh$.
Employing a proper order counting, we compute the total and
differential cross sections at the leading order, which
incorporate the one-loop corrections between zero external momenta
and their physical values.
Discovery/exclusion potential of a future $e^+e^-$ collider for this model is estimated.
We also find a unique feature in the momentum dependence of the Higgs triple vertex for this class of models.

\end{abstract}

\begin{keyword}

%% keywords here, in the form: keyword \sep keyword

\PACS 
%% PACS codes here, in the form: \PACS code \sep code

%% MSC codes here, in the form: \MSC code \sep code
%% or \MSC[2008] code \sep code (2000 is the default)

\end{keyword}

\end{frontmatter}

%% \linenumbers

%% main text
%\section{}
%\label{}

\section{Introduction}

The properties of the Higgs boson, now being uncovered 
through measurements at the
LHC experiments, seem to point to an increasingly consistent
picture with
predictions of the
Standard Model (SM) of particle physics.
In particular, up to now measured interactions of the Higgs boson with
other SM particles agree well with the SM predictions
and no sign of significant deviation
has been observed
%%%%%
\cite{Aaboud:2017sjh, Khachatryan:2014qaa,  Aad:2015iha, CMS:2017vru}.
%%%%%

In view of the current status, it would be worth considering
a class of models beyond the SM, which possess Higgs portal
couplings and at the same time induce
the electroweak symmetry breaking via the Coleman--Weinberg (CW)
mechanism
%%%%%
\cite{Coleman:1973jx}.
%%%%%
On the one hand, in such models 
anomalies tend to be
suppressed in the interactions among the SM particles
other than the Higgs boson,
as well as  in the interactions of
the Higgs boson with other SM particles.
On the other hand, 
large anomalies are predicted in the Higgs self-interactions, 
since the global shape of the Higgs potential
[which is given by a $\phi^4 \log \phi$--type potential at 
the leading-order (LO)] 
significantly deviates from that of the SM.
Furthermore, this class of models are discussed within the context of
dark matter physics
%%%%%
\cite{Burgess:2000yq,Ghosh:2017fmr,Silveira:1985rk,McDonald:1993ex,Cline:2013gha,Guo:2010hq,Endo:2015nba,Ishiwata:2011aa,Heikinheimo:2013fta,Gabrielli:2013hma,Davoudiasl:2014pya}.
%%%%%

Testability of this class of models has been studied.
The CW-type potential predicts a universal value
of the three-point Higgs self-coupling from computation of the one-loop effective
potential, which is given by
$\lambda_{hhh}=5/3\times\lambda_{hhh}^{\rm SM}$
\cite{Dermisek:2013pta,Endo:2015ifa,Hashino:2015nxa}.
The large deviation from the SM value can be probed relatively easily
using the process $e^+e^-\to Zhh$ \cite{Gounaris:1979px} at a future linear collider
with the center-of-mass 
%(c.m.)\ 
energy around $500$~GeV \cite{Barklow:2017awn}.

There is, however, a caveat in these testability analyses.
The value of the above three-point coupling is determined at
the zero external momentum limit $p_i\to 0$, and this constant value
has been used to scale the tree-level $h^3$--vertex in
the analyses.
The CW mechanism is unique in that certain
one-loop contributions become comparable to tree-level contributions,
the very reason why it is called a radiative symmetry breaking mechanism.
This feature applies to the Higgs self-interactions, and
one needs to include a part of the
one-loop corrections even in the
LO analyses, if the proper order counting is respected.
According to this order counting 
the one-loop corrections between $p_i =0$ and $p_i \sim {\cal O}(m_t)$
become formally the same order as the three-point
coupling determined at
$p_i\to 0$.
% This feature applies to the Higgs self-interactions, and
% the one-loop corrections between $p_i =0$ and $p_i \sim {\cal O}(m_t)$
% become formally comparable in size to their values
% at $p_i =0$.
% Therefore, one needs to include
% appropriate one-loop corrections even in the
% LO analyses, if the proper order counting is respected.
Most pessimistically before explicit computation, one could be
worried if
the large deviation predicted at $p_i =0$ 
may even be almost canceled at the physical values of the external momenta.

In this paper we compute the total and differential cross sections for
$e^+e^-\to Zhh$ in a CW-type Higgs portal model,
at the LO of the proper order counting.
We take up the minimal model analyzed in 
%%%%%
ref.\cite{Endo:2015ifa}
which includes $N$ singlet scalar particles
%%%%%
 and
use the order counting developed in 
%%%%%
ref.\cite{Endo:2016koi}.
%%%%% 
This model has a
high predictability because of the small number of model
parameters.
Let us describe briefly the current status of this model.
It is known that this model can be tested by direct dark
matter searches
%%%%%
\cite{McKay:2017iur}.
%%%%%
Using the most recent bounds by the XENON1T
 experiment
%%%%%
\cite{Aprile:2017iyp},
%%%%%
the model is excluded in the
region $N\ge 2$, while the case $N=1$ is marginal.
Nevertheless, this test assumes that the reheating temperature of the Universe
exceeds the singlet mass scale, and that the dark matter is thermalized. Hence the model cannot be excluded
without this assumption, or alternatively,
we can put bounds on the reheating
temperature ($T \ll m_s$).
It was also pointed that this model may be tested using $WW$ scattering processes
in the future 
%\cite{}
%%%%%
\cite{Endo:2016koi}.
%%%%%

We show that in this model one-loop corrections induce non-trivial
kinematical dependences to the
$e^+e^-\to Zhh$ cross sections, which cannot be accounted for by
the constant scaling of the Higgs triple coupling.
The kinematical dependence of the $h^3$--vertex reflects characteristic features of the model. We also show a general feature valid for CW-type Higgs portal models with more general non-SM sectors.
%We can probe the details of the model using these kinematical dependences.
%Thus, our analysis can provide a good template for analyses of (non-minimal)
%CW-type Higgs portal models with more complicated non-SM sectors.

In Sec.~2 we describe our model and its order counting rule.
In Sec.~3 we define an effective Higgs triple coupling.
The total and differential cross sections for $e^+e^- \to Zhh$
are computed in Sec.~4.
Conclusion is given in Sec.~5.
In \ref{app:loop}, loop functions are defined.
In \ref{app:effac}, a relation between the $h^3$--vertex and the Higgs wave function renormalization is derived using derivative expansion of the effective potential.
\section{CSI model}

\subsection*{Lagrangian}

We consider a model, which has an extended Higgs sector with classical
scale invariance (CSI).  Throughout the paper we adopt the Landau
gauge and dimensional regularization with $d=4-2\epsilon$ space-time
dimensions.
The bare Lagrangian of the CSI model is given by
\begin{align}
{\cal L}^{\rm CSI}
=&\,
 	\left[
 		{\cal L}^\text{SM}
 	\right]_{\mu^2_\text{H}\rightarrow 0}
 	+\frac{1}{2}
(\partial_{\mu}{\vec{S}_B})^{2}
	-{\lambda^{(B)}_\text{HS}}\,
		(H_B^{\dagger}H_B)(\vec{S}_B\cdot\vec{S}_B)
	-\frac{\lambda_{S}^{(B)}}{4}(\vec{S}_B\cdot\vec{S}_B)^2\,.
\label{eq:our_model}
\end{align}
$\vec{S}=(S_1, \cdots, S_N)^T$ denotes a real scalar field,
which is a SM singlet and belongs to the $N$ representation of a global
$O(N)$ symmetry.  The above Lagrangian is invariant under the SM gauge
symmetry and the $O(N)$ symmetry and is perturbatively renormalizable.  
$H$ denotes the doublet
Higgs field.  
Subscripts or superscripts ``$B$" in
eq.\,(\ref{eq:our_model}) show that the corresponding fields or
couplings are the bare quantities.  
The Higgs
interaction terms relevant in our analysis are given by
\begin{align}
{\cal L}^{\rm CSI}_\text{H--int}
&=
-\mu^{2\epsilon}(\lambda_{\rm H}+\delta\lambda_{\rm H})(H^\dagger H)^2
-\mu^{2\epsilon}(\lambda_{\rm HS}+\delta\lambda_{\rm HS})
H^\dagger H\,S_iS_i \,.
\label{CSI-HiggsSector}
\end{align}
Here we have re-expressed the interaction terms by renormalized
quantities and counter-terms: $H$ and $S_i$ denote the renormalized
fields; $\lambda_{\rm H}$ and $\lambda_{\rm HS}$ represent the
renormalized coupling constants; the terms proportional to
$\delta\lambda_{\rm H}$ and $\delta\lambda_{\rm HS}$ represent the
counter-terms; $\mu$ denotes the renormalization scale.

The Higgs field acquires a
non-zero vacuum expectation value (VEV) via the CW mechanism, whereas the singlet
field does not
\cite{Endo:2015ifa}.  
The singlet particles become massive and degenerate (with mass $m_s
%%%%%
=\lambda_{\rm HS}^{1/2}v
%%%%%
$).
We expand the Higgs field about the VEV as $H=(G^+,
(v\mu^{-\epsilon}+h+iG^0)/\sqrt{2})^T$ and set $S_i=0%s_i
$, where $h$,
$G^0$ and $G^+$ represent the physical Higgs, neutral- and charged-NG
bosons, respectively; $v$ denotes the Higgs VEV.  Substituting them into
eq.\,(\ref{CSI-HiggsSector}), one obtains the Feynman rules
for the CSI model.  
% The roles of the Higgs quartic couplings turn out to be quite
% different between the CSI model and the SM.

\subsection*{Order counting: $\xi$ expansion}

According to ref.\cite{Endo:2016koi} we introduce
an auxiliary expansion parameter $\xi$ and rescale the parameters of
the model as follows:
\begin{align}
  \lambda_{\rm HS}
  \rightarrow
  \LO\, \lambda_{\rm HS} \, ,
~~~
  \lambda_{\rm H}
  \rightarrow
  \NLO\, \lambda_{\rm H} \, ,
~~~
  y_t
  \rightarrow
  \LO^{1/2}\,   y_t ,
~~~
\mbox{others}
  \rightarrow
{\cal O}(\xi^2)\,,
\label{eq:coupling-orders}
\end{align}
where $y_t=\sqrt{2}m_t/v$ denotes the top-quark Yukawa coupling. 
Then we expand each physical observable in series
expansion in $\xi$, and in the end we set $\xi=1$.  
If an
observable is given as $A(\xi)= \xi^{n}(a_0 + a_1\xi + a_2 \xi^2 +
\dots )$, we define the LO term of $A$ as $a_0$, the next-to-leading
order (NLO) term of $A$ as $a_1$, etc.  
From previous experiences
we expect the size of the effective expansion parameter 
in these series expansions
to be order 10--30\%, depending on the observables.
It follows from the
above counting that $m_h^2\sim{\cal O}(\NLO)$ and
$m_t^2, m_s^2\sim{\cal O}(\LO)$. 
The reason for assigning $\xi^2$ to
the other couplings (in particular to the
electroweak gauge couplings) will be made clear below.

\subsection*{Determination of parameters}

%%%%%
Following Sec.~II of 
ref.\cite{Endo:2016koi},
we can determine $\lambda_{\rm H}$ and $\lambda_{\rm HS}$
from the tadpole condition and the on-shell Higgs mass condition
in terms of VEV $v=246.6~{\rm GeV}$,
the Higgs mass $m_H=125.03\pm 0.27~{\rm  GeV}$~\cite{Aad:2014aba,Khachatryan:2014ira}
and the top-quark mass $m_t=173.34\pm 0.76~{\rm GeV}$~\cite{ATLAS:2014wva}.
$\lambda_{\rm H}$ and $\delta\lambda_{\rm H}$ depend on renormalization scheme; if the counter-terms are defined by the $\overline{\rm MS}$ scheme,
\bea
\lambda_{\rm H}&=&\frac{N\lambda_{\rm HS}}{16\pi^2} 
\left(
1-\ln[\lambda_{\rm HS}v^2/\mu^2]
\right)
-\frac{3y_t^4}{16\pi^2} 
\left(
1-\ln[y_t^4v^2/(2\mu^2)]
\right), \\
\delta\lambda_{\rm H}&=&\frac{1}{\epsilon}\left(\frac{N\lambda_{\rm HS}}{16\pi^2}-\frac{3y_t^4}{16\pi^2}
\right).
\eea
The on-shell Higgs mass condition is given by
\be
\Sigma(m_h^2)=0, \label{eq:eq-of-lhs-det} %eq.\,(\ref{eq:eq-of-lhs-det}) 
\ee
where $\Sigma(q^2)$
denotes the Higgs self-energy, counted as\footnote{
Formally we can set 
$(q^2-m_h^2)/[q^2-m_h^2-\Sigma_h(q^2)]=1$
at LO in both regions $q^2\sim{\cal O}(m_h^2)$ and
$q^2\gg m_h^2$.
In the former region, we can expand $\Sigma_h(q^2)$
in $(q^2-m_h^2)/m_s^2\sim {\cal O}(\xi)$ and set 
$\Sigma_h(q^2)\to \Sigma_h(m_h^2)=0$;
in the latter region, $q^2-m_h^2\gg \Sigma_h(q^2)$
so that we can ignore $\Sigma_h(q^2)$.
Since, however, the hierarchy between $m_h^2\sim{\cal O}(\xi^2)$ 
and $m_t^2\sim{\cal O}(\xi)$ is not large, we prefer to keep this 
propagator ratio
in our computation.
}
 ${\cal O}(\xi^2)$, whose order is
the same as $m_h^2$.
The analytic expression of $\Sigma(q^2)$ is given in ref.\cite{Endo:2016koi}.
The values of $\lambda_{\rm HS}$ and $m_s$ determined by eq.\,(\ref{eq:eq-of-lhs-det}) and $v=246.6$GeV are summarized in Table~\ref{tab:lhsN1-2-12}.
%, $\lambda_{\rm HS}=4.87,2.45,1.42$ for $N=1,4,12$ respectively.
\begin{table}[tb]
  \begin{center}
    \begin{tabular}{|c|c|c|c|} \hline
$N$&$1$&$4$&12 \\ \hline \hline
$\lambda_{\rm HS}$ &$4.87$&$2.45$&$1.42$\\ \hline
$m_s~{\rm GeV}$&543 & 385&293 \\ \hline
    \end{tabular}
    \caption{Values of $\lambda_{\rm HS}$ and $m_s$ for $N=1,~4,~12$}
    \label{tab:lhsN1-2-12} % Table \ref{tab:lhsN1-2-12}
  \end{center}
\end{table}

%%%%%

\section{Higgs triple coupling} \label{sec:3}

\begin{figure}[tbp]
\begin{center}
\includegraphics[width=0.4\linewidth]{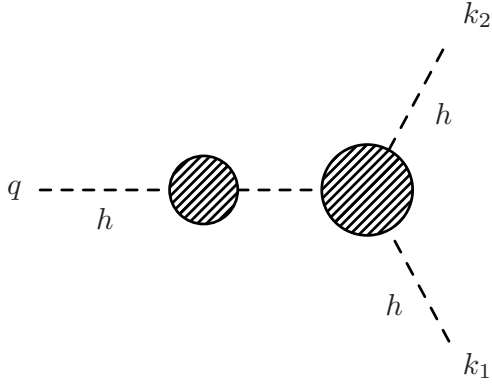}
\vspace*{-2mm}
 \caption{\small{Schematic diagram for the effective Higgs triple
coupling for $h^*\to hh$.}} 
 \label{EffTripleCoupling}
 \vspace*{-6mm}
\end{center} 
 \end{figure}

We define an effective Higgs triple
coupling for $h^*\to hh$ as
\bea
\lambda_{hhh}(q^2)=
\frac{q^2-m_h^2}{q^2-m_h^2-\Sigma_h(q^2)}\times
\frac{1}{v}\Gamma_{hhh}(q^2), \label{eq:triplecoup}
\eea
%%%%%
where $\Gamma_{hhh}(q^2)$
denotes the 1PI 
three-point vertex of the Higgs boson,
see Fig.~\ref{EffTripleCoupling}.
%%%%%
The two external Higgs bosons corresponding to the
final state are taken to be on-shell, while the
invariant mass $q$ of the initial (off-shell)
Higgs boson is taken as a variable.

\begin{figure}[t]
\begin{center}
    \begin{tabular}{cc}
      %---- ul
      \begin{minipage}[t]{0.4\hsize}
        \centering
        \includegraphics[keepaspectratio, scale=0.24]{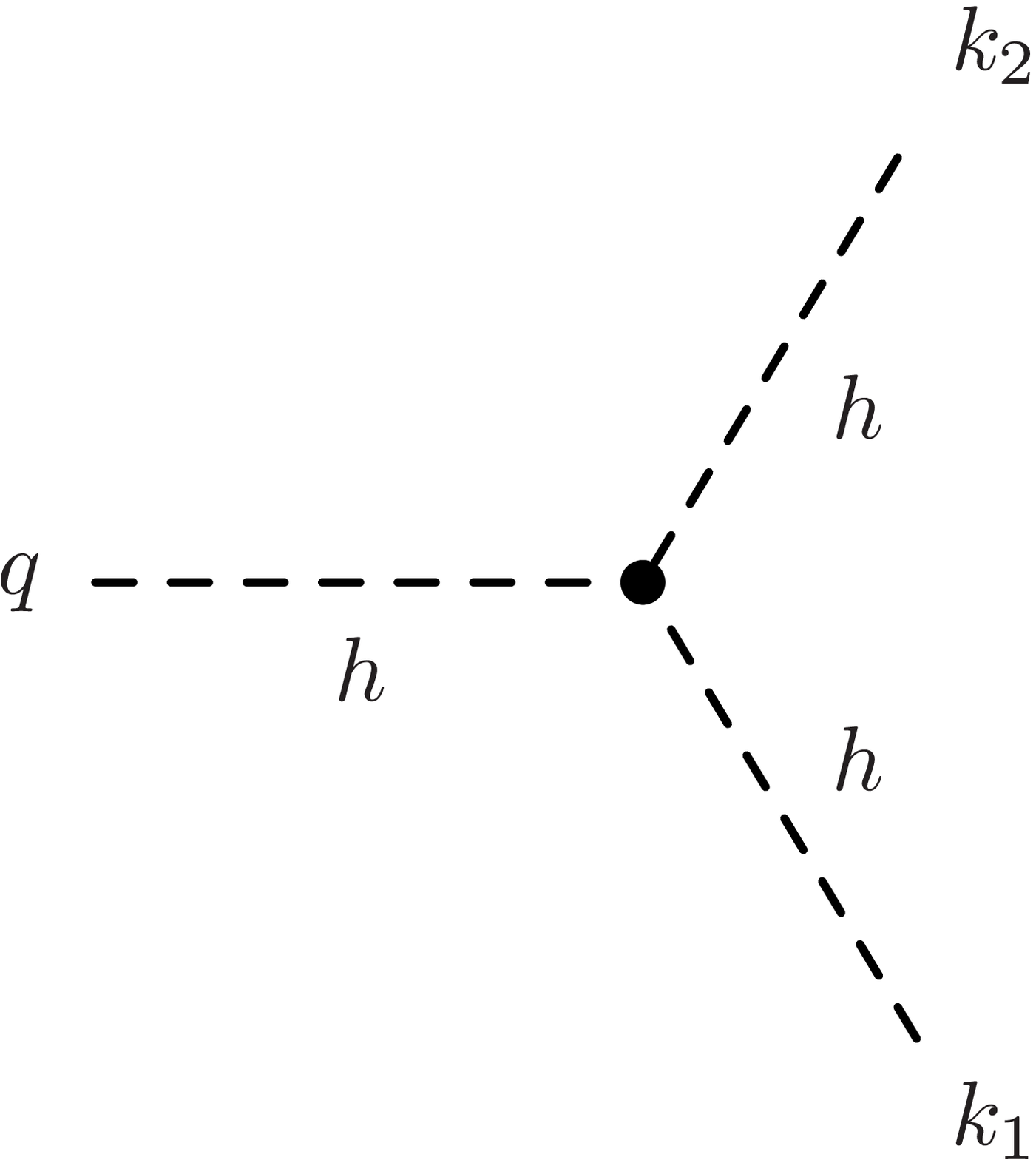}
      \end{minipage} &
      %---- ur
      \begin{minipage}[t]{0.4\hsize}
        \centering
        \includegraphics[keepaspectratio, scale=0.24]{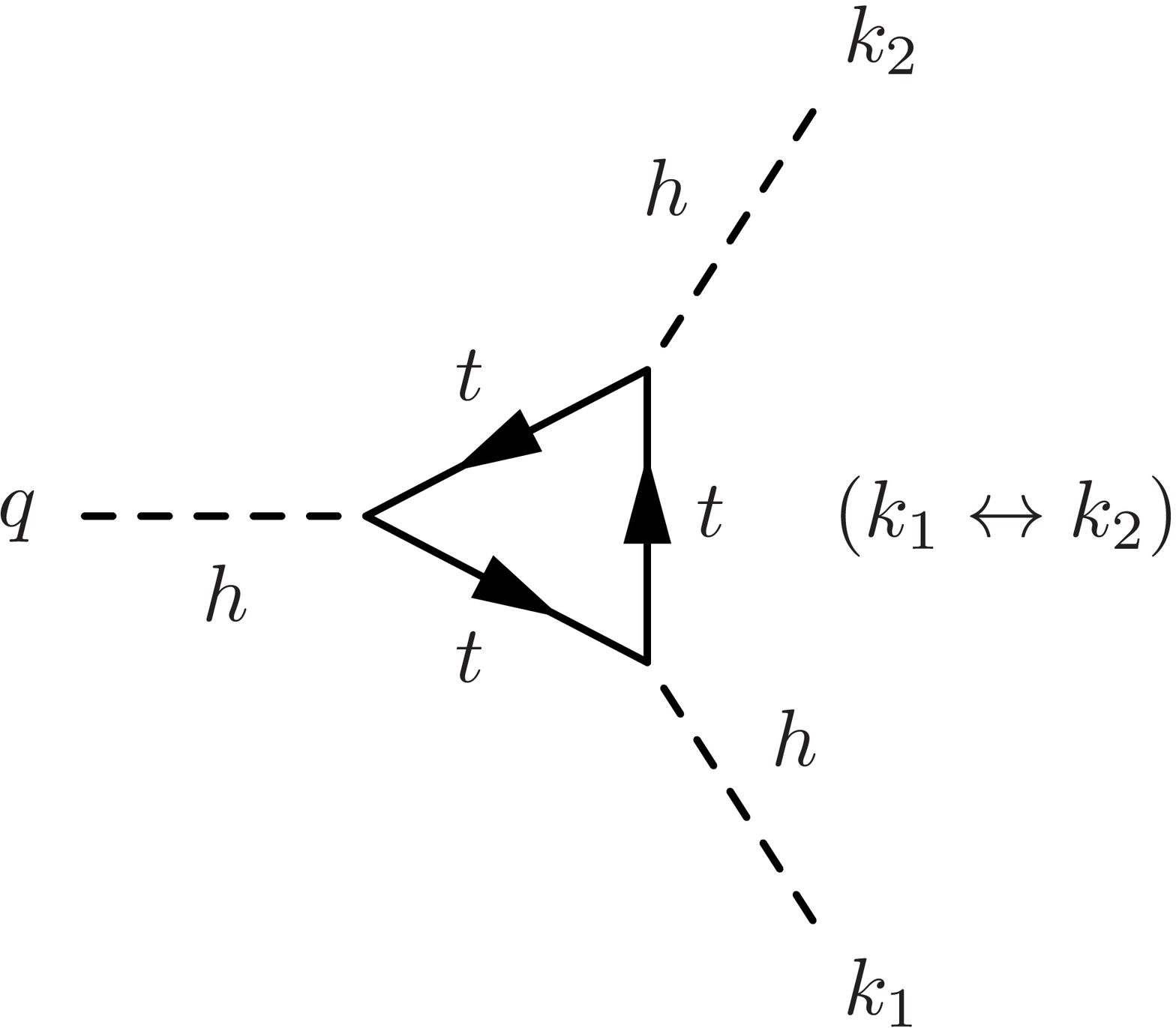}
      \end{minipage} \\
      %---- dl
      \begin{minipage}[t]{0.4\hsize}
        \centering
        \includegraphics[keepaspectratio, scale=0.24]{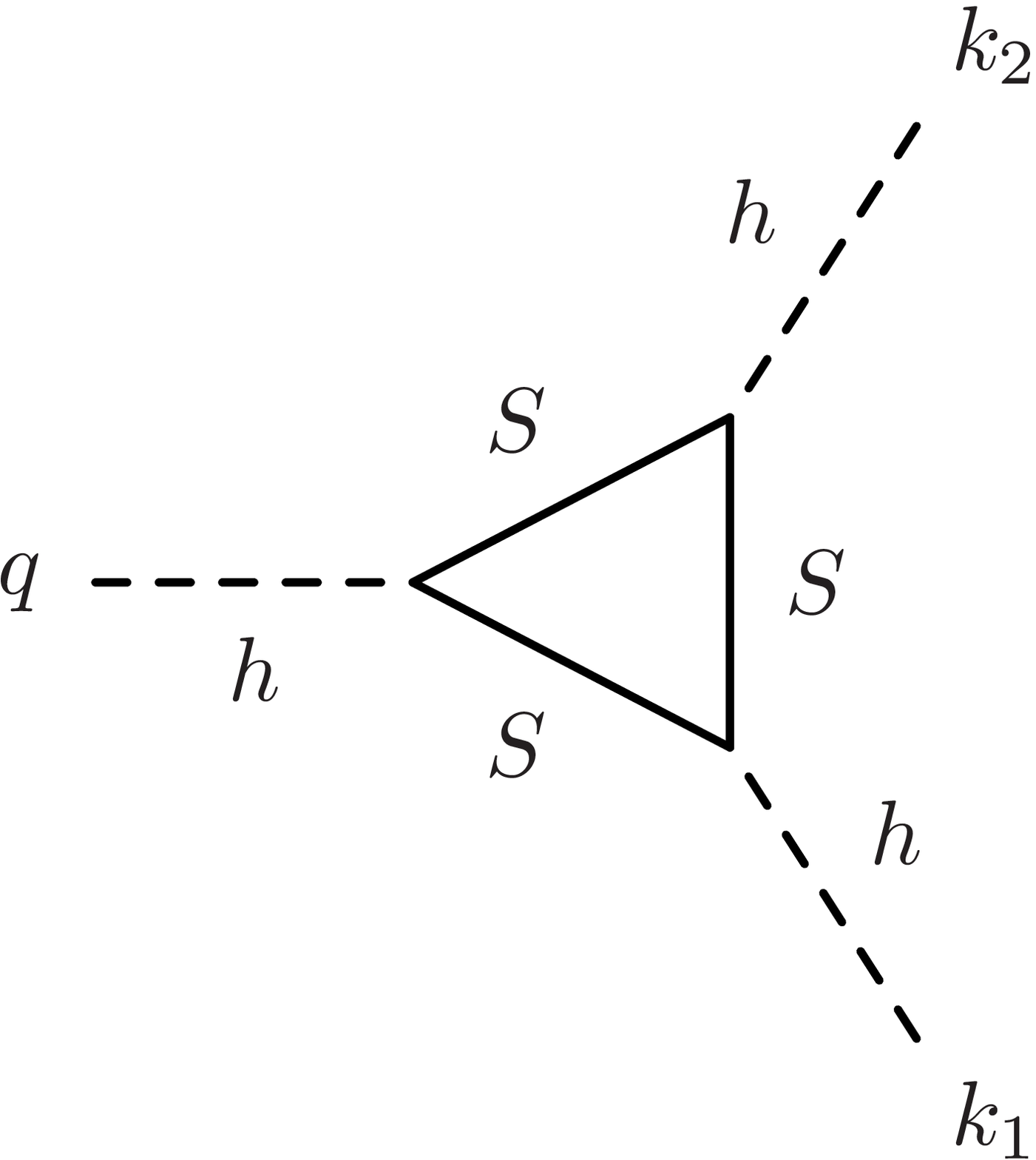}
      \end{minipage} &
      %---- dr
      \begin{minipage}[t]{0.4\hsize}
        \centering
        \includegraphics[keepaspectratio, scale=0.24]{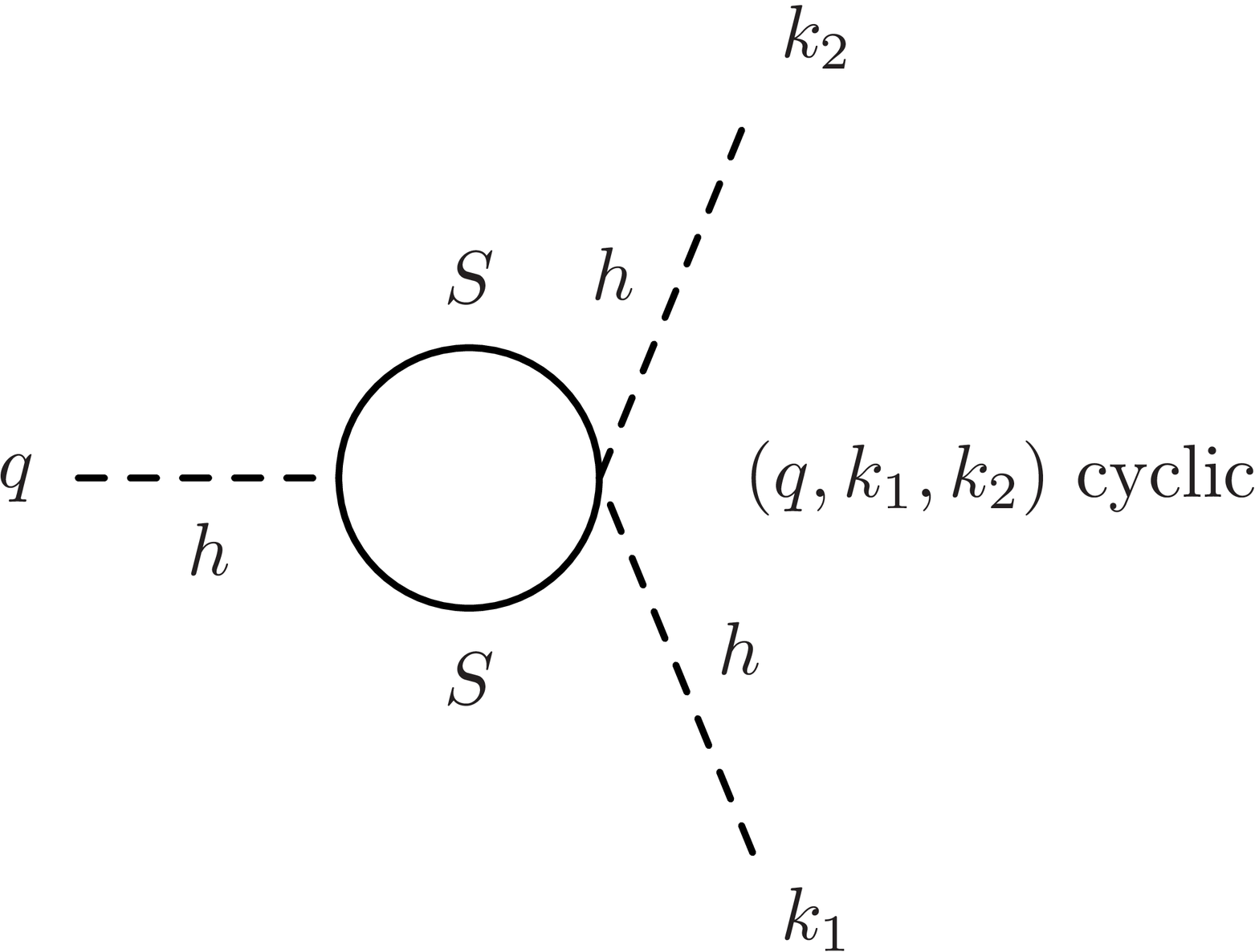}
      \end{minipage} 
    \end{tabular}
\vspace*{-2mm}
 \caption{\small{Diagrams contributing to $\Gamma_{hhh}(q^2)$ at LO.}} 
 \label{DiagramsTripleVertex}
 \vspace*{-6mm}
\end{center} 
 \end{figure}

%$\Sigma(q^2)$ is counted as\footnote{
%Formally we can set 
%$(q^2-m_h^2)/[q^2-m_h^2-\Sigma_h(q^2)]=1$
%at LO in both regions $q^2\sim{\cal O}(m_h^2)$ and
%$q^2\gg m_h^2$.
%In the former region, we can expand $\Sigma_h(q^2)$
%in $(q^2-m_h^2)/m_s^2\sim {\cal O}(\xi)$ and set 
%$\Sigma_h(q^2)\to \Sigma_h(m_h^2)=0$;
%in the latter region, $q^2-m_h^2\gg \Sigma_h(q^2)$
%so that we can ignore $\Sigma_h(q^2)$.
%Since, however, the hierarchy between $m_h^2\sim{\cal O}(\xi^2)$ 
%and $m_t^2\sim{\cal O}(\xi)$ is not large, we prefer to keep this 
%propagator ratio
%in our computation.
%}
%
% ${\cal O}(\xi^2)$, which is
%the same as $m_h^2$.
%Its analytic expression is given in \cite{Endo:2015nba} .
%
The diagrams contributing to $\Gamma_{hhh}(q^2)$
are shown in Fig.~\ref{DiagramsTripleVertex}.
Its analytic expression is given by
\bea
%m_h^2 +\Sigma(q^2)&=&
% 3(\lambda _H+\delta \lambda_H) v^2 \nonumber \\&&
% -\frac{3 y_t^2 }{16 \pi^2}(q^2-4m_t^2)B_0\left(q^2,m_t^2,m_t^2\right)
%+\frac{3 y_t^2 }{8 \pi ^2}A_0\left(m_t^2\right) \nonumber \\&&
%-\frac{ N \lambda _{\text{HS}}^2 v^2 }{8 \pi ^2} B_0\left(q^2,m_s^2,m_s^2\right)
%-\frac{N \lambda _{\text{HS}} }{16 \pi^2}A_0\left(m_s^2\right) \\
\frac{1}{v} \Gamma_{hhh} (q^2)&=&
6 (\lambda _H+\delta \lambda_H) \nonumber \\ &&
+\frac{3 m_t y_t^3 }{4 \sqrt{2} \pi ^2 v}B_0\left(q^2,m_t^2,m_t^2\right)
+\frac{3 m_t y_t^3 }{2\sqrt{2} \pi ^2 v}B_0\left(0,m_t^2,m_t^2\right) \nonumber \\ &&
-\frac{3  m_t y_t^3 }{8\sqrt{2} \pi ^2 v}(q^2-8m_t^2)C_0\left(q^2,0,0,m_t^2,m_t^2,m_t^2\right) \nonumber
\\&&
-\frac{N \lambda _{\text{HS}}^2 }{8 \pi ^2}B_0\left(q^2,m_s^2,m_s^2\right)
-\frac{N \lambda _{\text{HS}}^2 }{4\pi ^2}B_0\left(0,m_s^2,m_s^2\right) \nonumber \\&&
-\frac{N \lambda _{\text{HS} }^3 v^2 }{2 \pi ^2}C_0\left(q^2,0,0,m_s^2,m_s^2,m_s^2\right), \label{eq:gammaexpr}
\eea
where $B_0\left(p^2,m^2,m^2\right)$ and $C_0\left(p_1^2,p^2_2,p^2_3,m^2,m^2,m^2\right)$ represent loop functions defined in \ref{app:loop}.  $\Sigma(q^2)$ and $\Gamma(q^2)$ do not depend on the renormalization scale $\mu^2$, due to cancellation between $\lambda_{\rm H}$ and loop functions.
We can use $\lambda_{hhh}(q^2)/\lambda_{hhh}^{\rm SM,tree}$ to scale the
tree-level SM Higgs three-point vertex in the 
Higgs exchange diagram for $e^+e^-\to Zhh$.
\begin{figure}[tbp]
\begin{center}
%    \begin{tabular}{cc}
%      %---- l
%      \begin{minipage}[t]{0.4\hsize}
%        \centering
        \centering
        \includegraphics[keepaspectratio, scale=0.8] {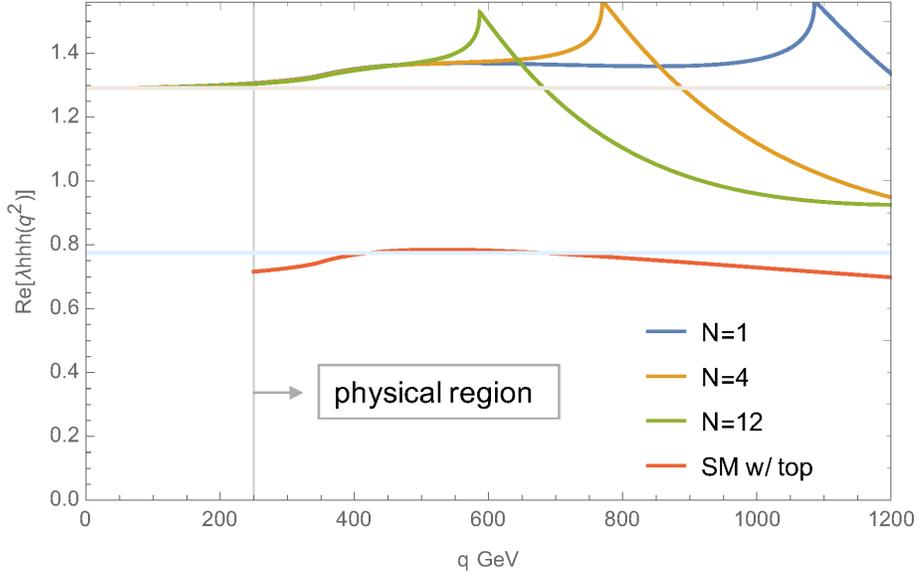}%{real.pdf}
\\(a)\\ 
%      \end{minipage}  
%      %---- r
%      \begin{minipage}[t]{0.4\hsize}
%        \centering
        \includegraphics[keepaspectratio, scale=0.8]{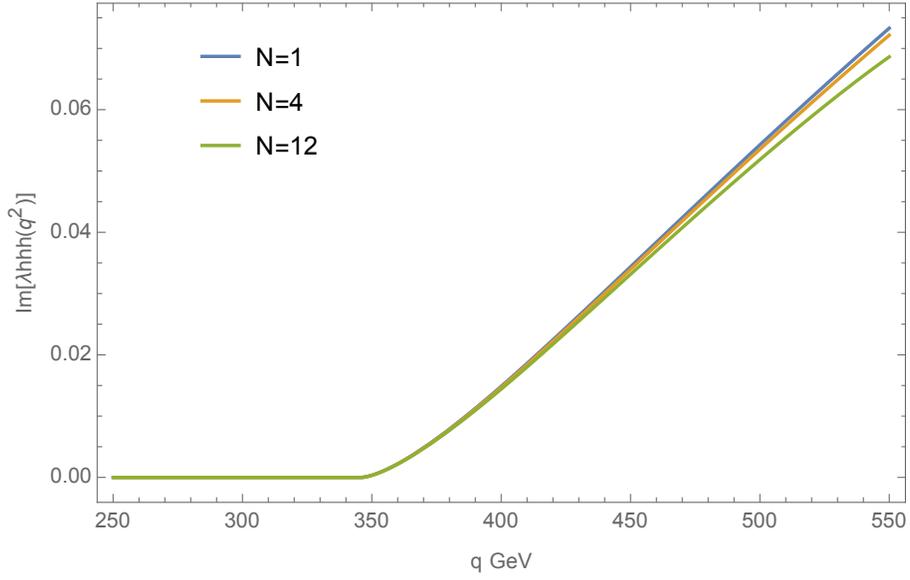}
\\(b)\\
%      \end{minipage} 
%    \end{tabular}
\vspace*{-2mm}
 %%%%%%
 \caption{\small{(a) ${\rm Re}\,\lambda_{hhh}(q^2)$ at LO (blue ($N=1$), orange ($N=4$), green ($N=12$)) vs.\ $q$. 
 Also shown are $\lambda_{hhh}$ as determined from the one-loop effective
 potential (light brown) and ${\rm Re}\,\lambda_{hhh}^{\rm SM}$ tree (light blue) and including the
 tree plus top-quark one-loop contributions (red).
Physical region, indicated by the arrow, corresponds to $2m_h\le q$. % \le \sqrt{s}-m_Z$.
(b) Same as (a) but for ${\rm Im}\,\lambda_{hhh}(q^2)$.}} 
%%%%%%
 \label{Plot-MomDep-EffCoupling} %Fig.~\ref{Plot-MomDep-EffCoupling}
 \vspace*{-6mm}
\end{center} 
 \end{figure}
We show the real and imaginary parts of
$\lambda_{hhh}(q^2)$ 
as functions of $q$ in 
Figs.~\ref{Plot-MomDep-EffCoupling} a,b.
%%%%%
We use eq. (\ref{eq:triplecoup}) to make the plot.
%%%%%
For comparison we also show
$\lambda_{hhh}$ as determined from the one-loop effective potential, $\frac{5}{3}\lambda^{\rm SM}_{hhh}$,
which corresponds to setting all the external momenta to zero.
By comparison we see non-trivial $q^2$ dependence in $\lambda_{hhh}(q^2)$.
In the same figures we also show
$\lambda_{hhh}^{\rm SM}$ including the
 tree plus top-quark one-loop contribution,
where the $q^2$ dependence stems from the top-loop contribution.
The top-quark loop contribution raises the couplings at
$q\simgt 2m_t$, which is common in the SM and CSI model.

%%%%%
It is useful to %see
examine
 $q^2$ expansion of $\lambda_{hhh}(q^2)$ for model identification.
$q^2$ expansion is reasonable for singlet contribution at $q\simeq 2m_h$, because $q^2/m_s^2 \ll 1$.
We show in \ref{app:effac}, using derivative expansion of the effective action, that the coefficient of the $q^2$ term of this expansion is determined by the divergent part of the wave function renormalization for the Higgs boson in a
general CSI-model.
Since the singlet-loop does not contribute to the divergent part of the Higgs wave function renormalization, there is no $q^2$ term from the singlet-loop in $\lambda_{hhh}(q^2)$.
The absence of $q^2$ term in the singlet contribution is also confirmed by an explicit calculation of $\Gamma_{hhh}$:
\bea
&&-\frac{N \lambda _{\text{HS}}^2 }{8 \pi ^2}B_0\left(q^2,m_s^2,m_s^2\right)
-\frac{N \lambda _{\text{HS}}^2 }{4\pi ^2}B_0\left(0,m_s^2,m_s^2\right)
-\frac{N \lambda _{\text{HS} }^3 v^2}{2 \pi ^2}C_0\left(q^2,0,0,m_s^2,m_s^2,m_s^2\right) \nonumber \\
&=&({\rm const.})
+\left[\frac{1}{6}\frac{N \lambda _{\text{HS}}^2 }{8\pi ^2}
-\frac{1}{24m_s^2}\frac{N \lambda _{\text{HS} }^3 v^2}{2 \pi ^2}
\right]\left(  \frac{q^2}{m_s^2}  \right)
+\left[\frac{1}{60}\frac{N \lambda _{\text{HS}}^2 }{4\pi ^2}
-\frac{1}{180m_s^2}\frac{N \lambda _{\text{HS} }^3 v^2}{2 \pi ^2}
\right]\left(  \frac{q^2}{m_s^2}  \right)^2 +\cdots \nonumber \\
&=&({\rm const.})
+0\times \frac{q^2}{m_s^2}
+\frac{N\lambda^2_{\rm HS}}{1440\pi^2}\left(  \frac{q^2}{m_s^2}  \right)^2+ \cdots. \label{eq:cancel}
\eea
See eq.(\ref{eq:gammaexpr}).
In the third line, $m_s^2=\lambda_{\rm HS} v^2$ is used. 
Thus, the $q^2$ dependence starts at order $q^4$ for the singlet contribution.
On the other hand, since the top-quark contributes to the divergent part of the Higgs wave function renormalization, the $q^2$ term of $\lambda_{hhh}(q^2)$ stems solely from the top-quark contribution at LO.
This is why $N$ dependence is almost absent until very close to the singlet-pair threshold in Fig.~\ref{Plot-MomDep-EffCoupling} a,b.
%%%%%
Hence, using this unique feature of CSI models, we can test the origin if the anomaly by looking at the behavior of the Higgs triple coupling near the Higgs pair threshold.
%%%%%

\section{$e^+e^- \to Zhh$ cross sections}

\begin{figure}[tb]
\begin{center}
    \begin{tabular}{cc}
      %---- ul
      \begin{minipage}[t]{0.45\hsize}
        \centering
        \includegraphics[keepaspectratio, scale=0.2]{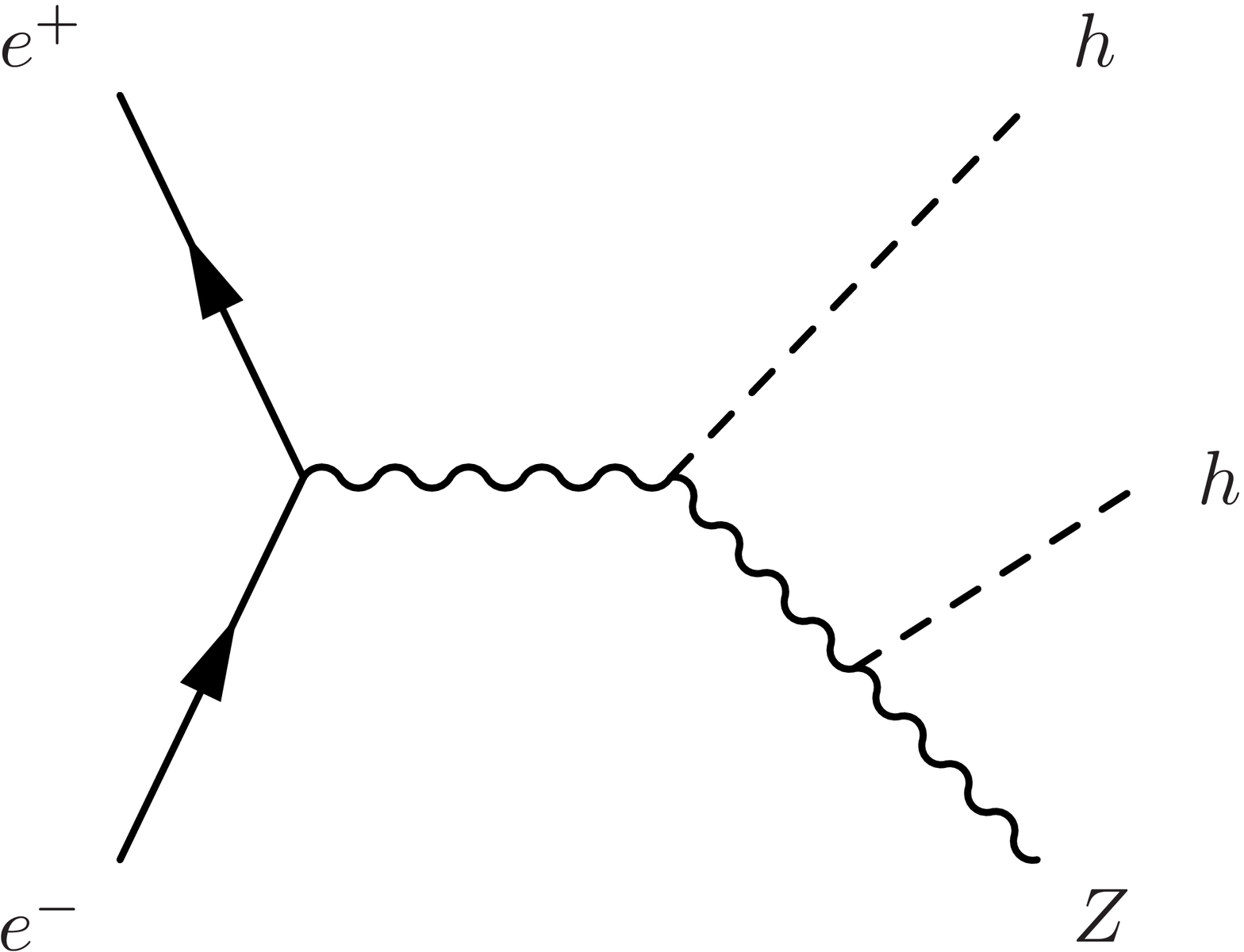}
      \end{minipage} &
      %---- ur
      \begin{minipage}[t]{0.45\hsize}
        \centering
        \includegraphics[keepaspectratio, scale=0.2]{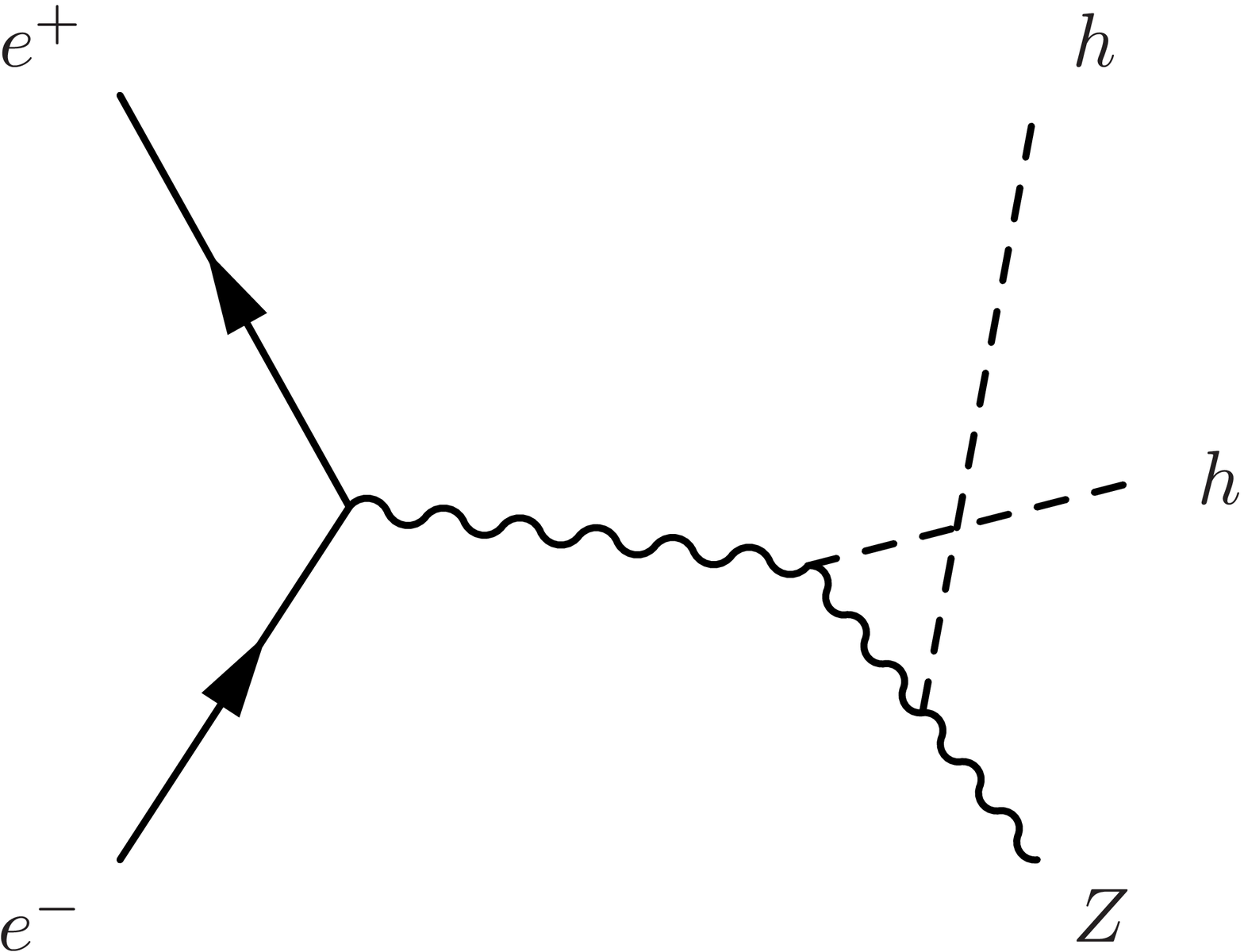}
      \end{minipage} \\
      %---- dl
      \begin{minipage}[t]{0.45\hsize}
        \centering

        \includegraphics[keepaspectratio, scale=0.2]{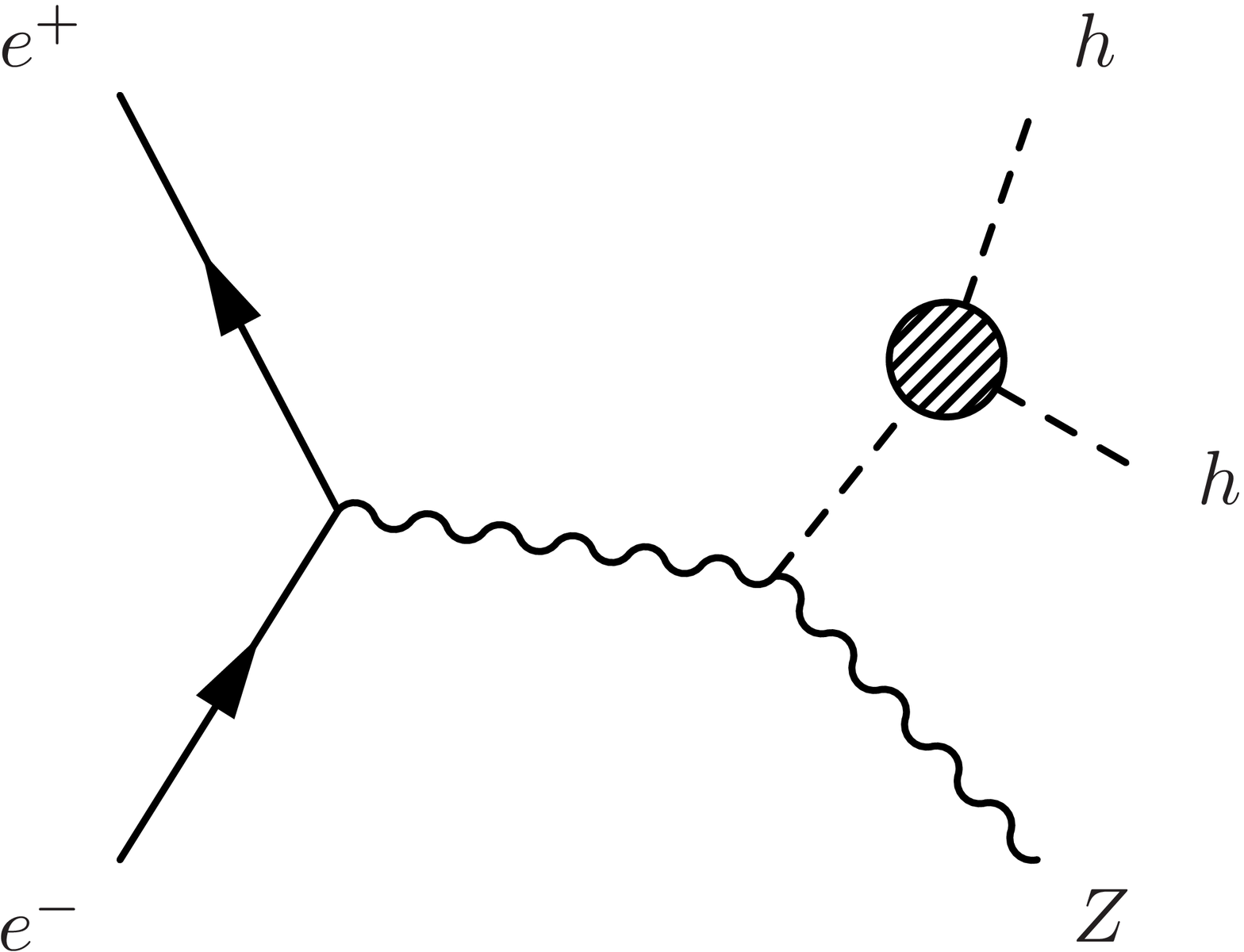}
      \end{minipage} &
      %---- dr
      \begin{minipage}[t]{0.45\hsize}
        \centering
        \includegraphics[keepaspectratio, scale=0.2]{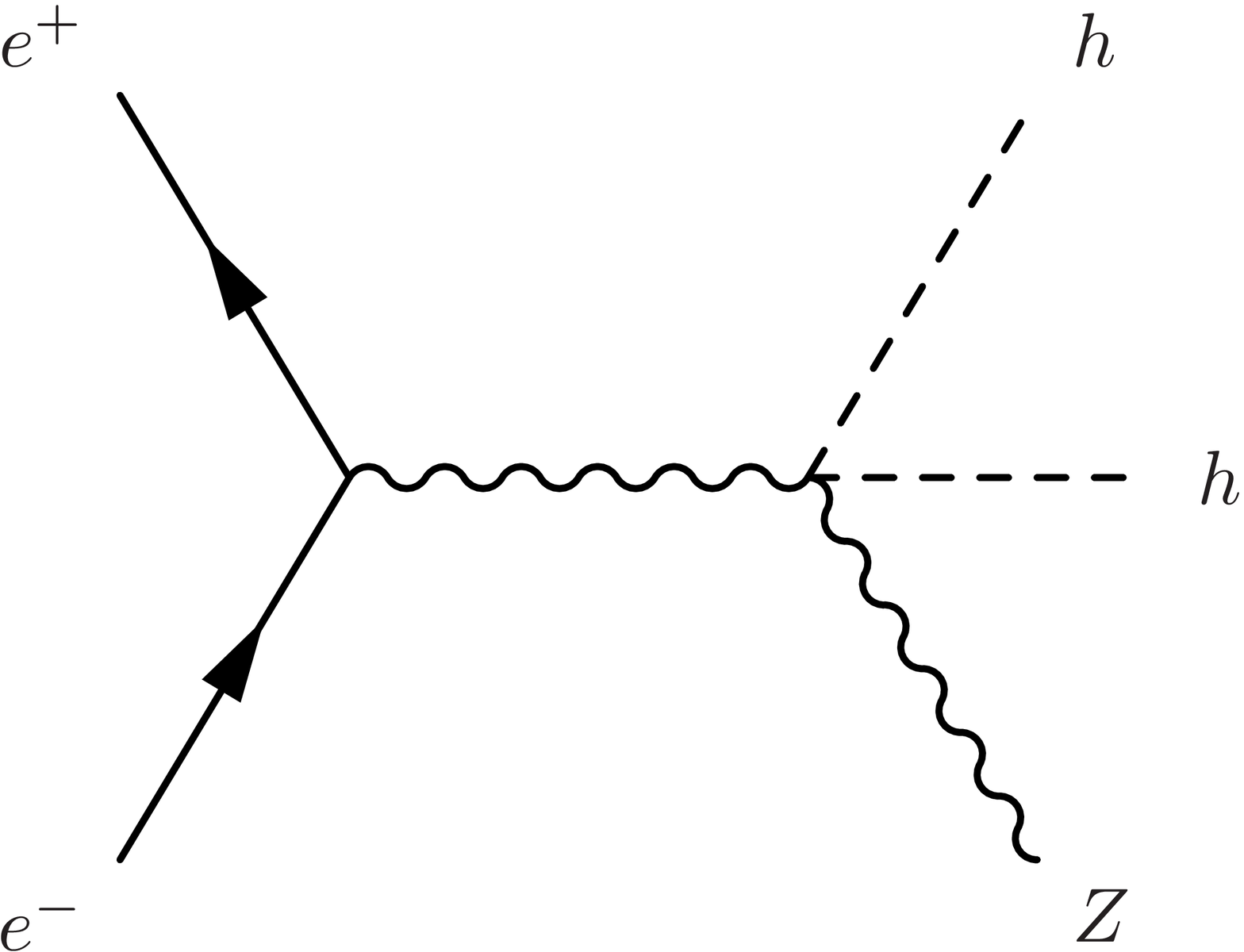}
      \end{minipage} 
    \end{tabular}
\vspace*{-2mm}
 \caption{\small{
Diagrams which contribute to $e^+e^- \to Zhh$.
}} 
 \label{Diagrams-e+e-ToZhh}
 \vspace*{-6mm}
\end{center} 
 \end{figure}

In the SM there are four tree-level diagrams which contribute to the
process $e^+e^- \to Zhh$.
See Fig.~\ref{Diagrams-e+e-ToZhh}.
One of the diagrams contains the Higgs three-point vertex,
while the other three diagrams contribute as irreducible
background diagrams for probing the triple Higgs coupling.
To compute the cross section for the CSI model, 
we replace the tree-level $\lambda_{hhh}^{\rm SM}$
by $\lambda_{hhh}(q^2)$.
The background diagrams and the signal diagram are counted
as the same order in $\xi$ since we assign
$\xi^2$ to the electroweak gauge couplings.
Numerically this order counting is reasonable.

\begin{figure}[tbp]
\begin{center}
\includegraphics[width=0.7\linewidth]{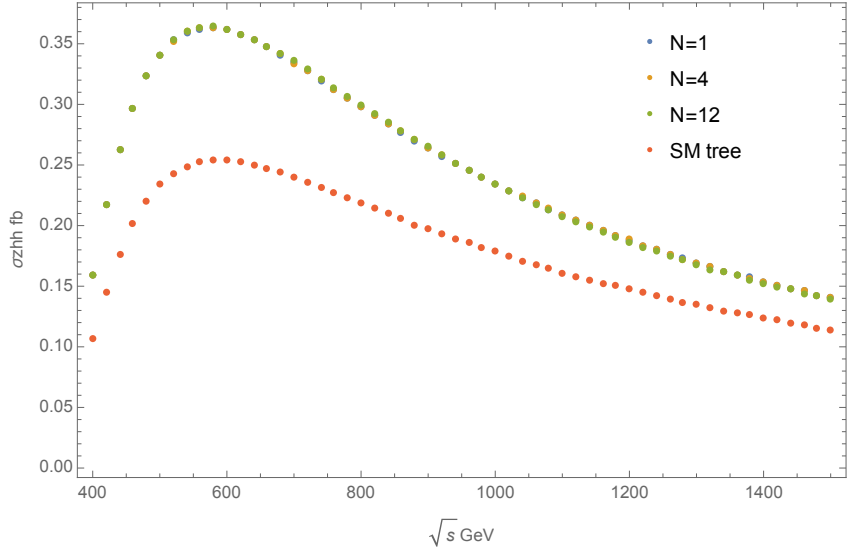}
\vspace*{-2mm}
 \caption{\small{
Total cross section $\sigma$ for $e^+e^-\to Zhh$
vs. $\sqrt{s}$
 at SM (red), $N=1$ (blue), $N=4$ (orange),$N=12$ (green).
}} 
 \label{Xsecvssqrts} %Fig.~\ref{Xsecvssqrts}
 \vspace*{-6mm}
\end{center} 
 \end{figure}

\begin{figure}[tb]
\begin{center}
\includegraphics[width=0.7\linewidth]{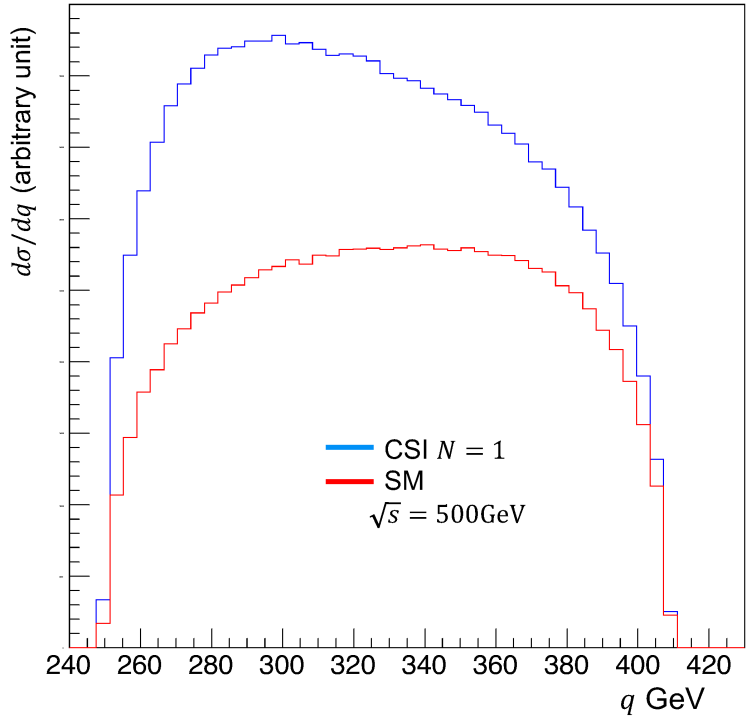}
\vspace*{-2mm}
 \caption{\small{
 Differential cross section
$d\sigma/dq$ at $\sqrt{s}=500$~GeV for $e^+e^- \to Zhh$, where
$q^2=(p_{h1}+p_{h2})^2$. 
$d\sigma/dq$ is non-zero only in $2m_h\le q \le \sqrt{s}-m_Z$. 
%The difference between CSI (blue) and SM (red) mainly appears in low $q$ region, since the signal diagram is suppressed by the Higgs propagator at large $q$ region.
}} 
 \label{Histo-DifferentialXsec}
 \vspace*{-6mm}
\end{center} 
 \end{figure}

\begin{figure}[tb]
\begin{center}
\includegraphics[width=0.7\linewidth]{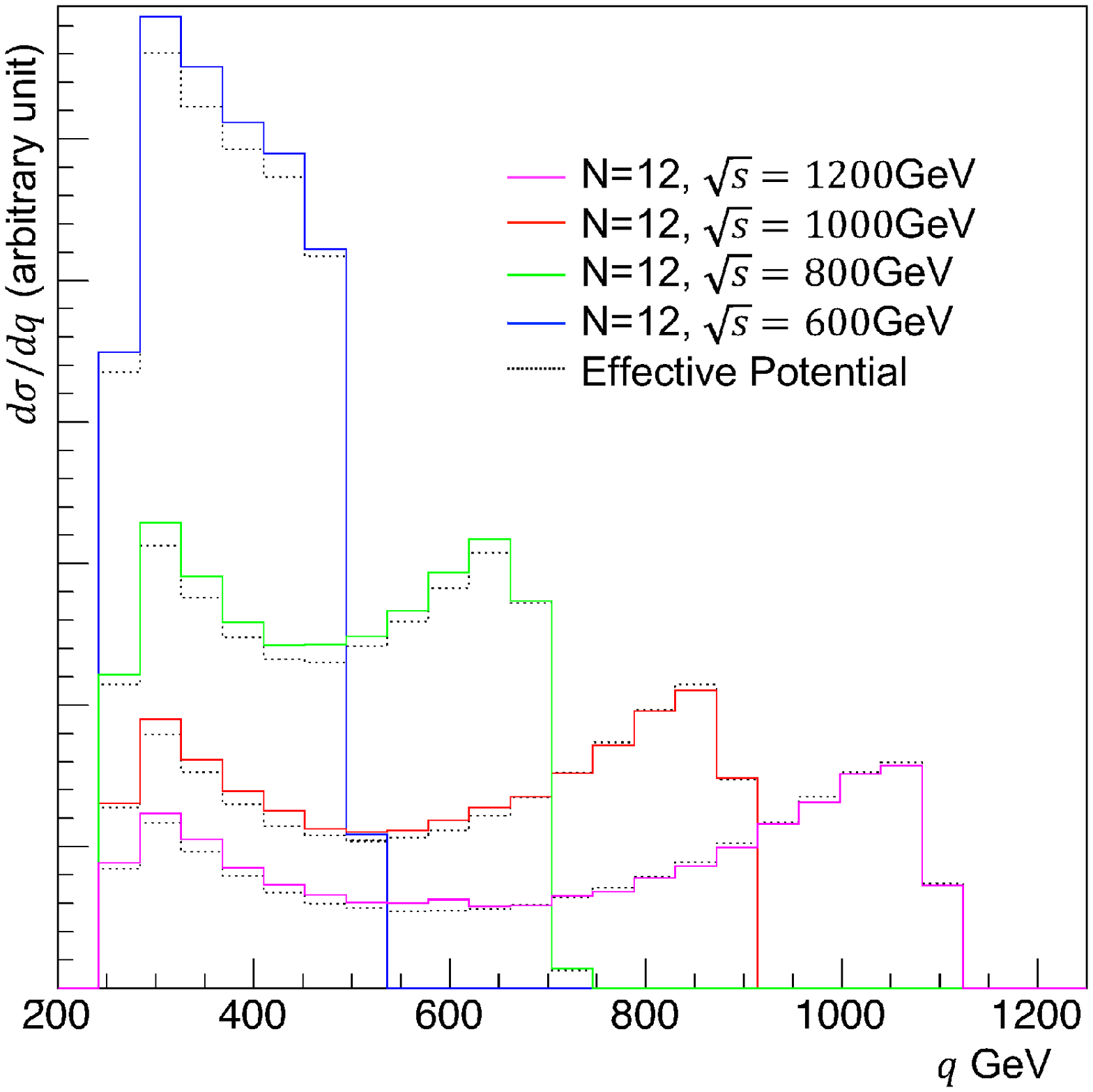}
\vspace*{-2mm}
 \caption{
 Differential cross sections \small{$d\sigma/dq$ at $N=12$, $\sqrt{s}=600({\rm blue})$, $800({\rm green})$, $1000({\rm red})$ and $1200({\rm magenta})$ GeV for $e^+e^- \to Zhh$, where
$q^2=(p_{h1}+p_{h2})^2$. Black dotted lines show the same cross sections using $\lambda_{hhh}$ determined from the effective potential. 
%Colored histogram should have the singlet pair creation peak in $q \simeq 2m_s\simeq 600$ GeV, but there is no peak since the signal diagram is suppressed by the Higgs propagator.
}} 
 \label{Histo-diffxsecN12} %Fig.~\ref{Histo-diffxsecN12}
 \vspace*{-6mm}
\end{center} 
 \end{figure}

%We calculate the total and differential
%cross sections using MadGraph5
%, with
%the initial $e^\pm$ longitudinal polarizations
%$P(e^+,e^-)=(0.3,-0.8)$.
%At $\sqrt{s}=500$~GeV, the total cross section
%is evaluated to be $\sigma^{\rm CSI}_{\rm tot}(e^+e^- \to Zhh)=0.341$~fb.
%This amounts to $+47$\% deviation compared to the tree-level
%SM total cross section.
 %%%%%
We calculate the total and differential
cross sections using MadGraph5 \cite{Alwall:2014hca}, with
the initial $e^\pm$ longitudinal polarizations
$P(e^+,e^-)=(0.3,-0.8)$.
At $\sqrt{s}=500$~GeV, the total cross section
is evaluated 
to be $\sigma^{\rm CSI}_{\rm tot}(e^+e^- \to Zhh)=0.341$~fb.
This amounts to $+47$\% deviation compared to the tree-level
SM total cross section. We show the $\sqrt{s}$ dependence of the total cross section in
Fig.~\ref{Xsecvssqrts}.
The deviation of the total cross section from the tree-level SM prediction decreases as $\sqrt{s}$ increases.
%%%%%
%{\it The plot for $\sqrt{s}$ dependence of the total cross sections
%is necessary.}
%We also compute $d\sigma/dq$ at
% $\sqrt{s}=500$~GeV , where
%$q^2=m_{hh}^2=(p_{1}+p_{2})^2$.
%The result is shown in Fig.~\ref{Histo-DifferentialXsec}.
%%%%%
We also compute $d\sigma/dq$ at
 $\sqrt{s}=500$~GeV for $N=1$ and at several $\sqrt{s}$ between $600-1200$~GeV for $N=12$, where
$q^2=m_{hh}^2=(p_{1}+p_{2})^2$.
The results are shown in Figs.~\ref{Histo-DifferentialXsec} and \ref{Histo-diffxsecN12}.
%%%%%
%The kinematically allowed range is given by $2m_h\le q \le \sqrt{s}-m_Z$.
%The enhancement from the SM prediction
%in the low $q$ region is due to the enhancement of the
%triple Higgs coupling at low $q$.
%The relative enhancement factor decreases as $q$ increases, since
%the relative weight of the signal diagram is reduced due to
%rapid decrease of the Higgs propagator $1/(q^2-m_h^2-\Sigma_h)$.
%%%%%
The kinematically allowed range is given by $2m_h\le q \le \sqrt{s}-m_Z$.
In Fig.~\ref{Histo-DifferentialXsec}, the enhancement from the SM prediction
in the low $q$ region is due to the enhancement of the
triple Higgs coupling at low $q$.
The relative enhancement factor decreases as $q$ increases, since
the relative weight of the signal diagram is reduced due to
rapid decrease of the Higgs propagator $1/(q^2-m_h^2-\Sigma_h)$.
In Fig.~\ref{Histo-diffxsecN12}, no peak is visible corresponding to the singlet pair creation at $q\simeq 2m_s (\simeq 600{\rm GeV})$, because of the suppression by  the Higgs propagator. The difference between the prediction using $\lambda_{hhh}(q^2)$ and $\lambda_{hhh}=\frac{5}{3}\lambda^{\rm SM}_{hhh}$ determined from the effective potential, is found in small $q$ region in the $N=12$ case.
%%%%%
From these figures we see that it is challenging to detect different $q$-dependences, which would require huge statistics.
%%%%%
% It is consistent with the region contributed by signal diagram.
%%%%%
%
%From detailed examination of
%the $q^2$ dependence, one can extract the $q^2$ dependence
%of $\lambda_{hhh}(q^2)$.
%
%{\it How is the $N$ dependence?}

%{\it Luminosity estimate}
%%%%%

%%%%%
Integrated luminosity necessary for a discovery at $5\sigma$ (an exclusion at $3\sigma$) of this model is estimated to be $710 ~{\rm fb}^{-1}$ ($260~ {\rm fb}^{-1}$) at $\sqrt{s}=500 {\rm GeV}$.
This estimation uses the number of signal and background events from the full simulation of International Linear Collider (ILC) experiment given in ref.\cite{Tian:2013qmi} for $e^+e^-\to Zhh$, $h\to b\bar{b}$ process and $m_h=120$GeV at $\sqrt{s}=500$GeV, with events corresponding to $2{\rm ab}^{-1}$.
We rescaled the number of the signal event 
$N_{\rm sig}$ by the ratio of the total cross sections and of the branching ratios for $h\to b\bar{b}$ as
\be
N_{\rm sig}\to N^{\rm CSI}_{\rm sig}=(\sigma^{\rm CSI}_{\rm tot}/\sigma^{\rm SM}_{\rm tot})(Br^{m_h=125{\rm GeV}}_{h\to b\bar{b}}/Br^{m_h=120{\rm GeV}}_{h\to b\bar{b}})^2N_{\rm sig},
\ee
while the $1\sigma$ standard deviation is approximated by $N_{\rm sig}/\sqrt{N_{\rm BG}}$. 
%%%%%

To end this section we give some discussion. The $W$-fusion process is another process at ILC to measure the triple Higgs coupling and is the dominant process at $\sqrt{s}\gtrsim 1200$GeV in SM. It turns out that the total cross section of this process decreases as the triple Higgs coupling increases due to a negative interference \cite{Tian:2013yda}, while the $Zhh$ cross section are enhanced by positive interference. As a result it is advantageous to analyze $e^+e^-\to Zhh$ rather than $W$-fusion process for the CSI model.
%{\it Discussion on Landau pole.}
%%%%%
\par
It is pointed that this model has a Landau pole around $3.5 ~{\rm TeV}~(N=1)$, $16 ~{\rm TeV}~(N=4)$ and $28 ~{\rm TeV}~(N=12)$ at LO \cite{Endo:2015ifa}. 
The existence of the Landau pole indicates that perturbative expansion of the model does not work around its scale or the model turns into some UV theory. 
Perturbative validity has been discussed in the leading-logarithmic analyses of the effective potential and $WW$-scattering amplitude \cite{Endo:2016koi}.  
Since our analysis deals with the energy scale well below the Landau pole, we consider that the perturbative analysis given in this paper is justified, where we regard this model as an effective theory valid around $\mathcal{O}(100 {\rm GeV})$.
%perturbation
%->LL-resum of eff pot and WW-scatt
%->CSI as a effective theory around EW scale

%\section{Results}

\section{Conclusion}

In the minimal CSI model,
we have incorporated the one-loop corrections between
the zero external momenta and physical point in the total and differential
cross sections for $e^+e^-\to Zhh$.
We find that the bulk of the large anomaly predicted at
the zero-external-momentum limit remains,
while a non-trivial $q^2$ dependence of the
Higgs triple interaction is induced.
%We can use the kinematical dependences of the cross sections
%to probe the detailed structure of the model.
%%%%%
We also find that at LO, the effect of the singlet scalar boson has no $q^2$ term in $q^2$ expansion of the Higgs triple coupling.
This feature does not depend on the number or mass spectrum of the singlet scalars and is a general feature of CW-type models with singlet scalar bosons, as shown in \ref{app:effac}. 
%%%%%%%%%%
The top loop effects induce a non-negligible $q^2$ dependence, in accord with the expectation based on order counting. In contrast, $N$-dependent effects by the singlet-loop are found to remain small below the singlet pair threshold due to this unique feature of the $q^2$ expansion. We have estimated sensitivity of the $e^+e^-\to Zhh$ total cross section to the deviation from the SM and found that it is fairly promising.
%%%%%%%%%%

\section*{Acknowledgments}
The authors are grateful to H.~Yokoya for useful
advices for the simulation studies.
The works of Y.F.\ and Y.S.\ are supported in part by 
%{\it GPPU}
%%%%% 
Graduate Program on Physics for the Universe (GP-PU),  Tohoku University,
%%%%%
and by Grant-in-Aid for
scientific research (No.~17K05404) from
MEXT, Japan, respectively.

%% The Appendices part is started with the command \appendix;
%% appendix sections are then done as normal sections
%% \appendix

%% \section{}
%% \label{}

\appendix
\section{Loop functions
%Passarino-Veltman intergrals
}\label{app:loop} %\ref{app:loop}
The loop functions used in Sec. \ref{sec:3} are defined as follows. Here, $d=4-2\epsilon$, $\mu$ represents the renormalization
scale, and $\kappa=(2\pi \mu)^{2\epsilon}/(i\pi^2)$:
\bea
%A_0(m^2)&=&\kappa \int \frac{d^dq}{q^2-m^2}, \\
B_0(p^2,m^2,m^2)&=&\kappa\int \frac{d^dq}{ [q^2-m^2][(q+p)^2-m^2]},\\
C_0(p_1^2,p_2^2,p_3^2,m^2,m^2,m^2)&=&\kappa \int \frac{d^dq}{ [q^2-m^2][(q+p_1)^2-m^2][(q+p_2)^2-m^2]}.
\eea
These can be expressed as
\bea
%A_0(m^2)&=&m^2 \left[\frac{1}{\bar{\epsilon}}+1-\ln(m^2/\mu^2)\right]
%, \\
B_0(p^2,m^2,m^2)&=&\frac{1}{\bar{\epsilon}}+\ln[\mu^2]-\int^1_0dx \ln[m^2-x(1-x)p^2-i0],
\eea
\be
C_0(p_1^2,p_2^2,p_3^2,m^2,m^2,m^2)=
\int^1_0dx \int^{1-x}_0 \!\!\!\!\!dy [m^2 -x(1-x)p_1^2+ xy(p^2_3-p^2_1-p^2_2)
-y(1-y)p_2^2-i0]^{-1},
\ee
where $p_1+p_2+p_3=0$, and $1/\bar{\epsilon}=1/\epsilon -\gamma+\ln[4\pi]$ ($\gamma=0.5772\cdots$ denotes Euler's number).\par
The expansion of $B_0(p^2,m^2,m^2)$ and $C_0(p^2,0,0,m^2,m^2,m^2)$ in $p^2$ are given by
\bea
B_0(p^2,m^2,m^2)&=&\frac{1}{\bar{\epsilon}}-\ln\left(  \frac{m^2}{\mu^2}  \right) -\frac{1}{6}\left(  \frac{p^2}{m^2}  \right)-\frac{1}{60}\left(  \frac{p^2}{m^2}  \right)^2 +\cdots,\\
C_0(p^2,0,0,m^2,m^2,m^2)&=&\frac{1}{2m^2}+\frac{1}{24m^2}\left(  \frac{p^2}{m^2}  \right) +\frac{1}{180m^2}\left(  \frac{p^2}{m^2}  \right)^2+\cdots.
\eea

\section{Expansion of effective action and momentum dependence of $\lambda_{hhh}(q^2)$} \label{app:effac}
We consider a general CSI model. The derivative expansion of 
the effective action in terms of classical Higgs field $\Phi$ in the symmetric phase is given by
\bea
\Gamma[\Phi]&=&-V(\Phi^{\dag}\Phi) \nonumber \\
&&+(\partial_{\mu} \Phi)^{\dag}(\partial_{\mu} \Phi )~\Gamma^{1,1}[\Phi^{\dag}\Phi]
+\Phi^{\dag}(\partial^2 \Phi )\Gamma^{1,2}[\Phi^{\dag}\Phi] \nonumber \\
&&+((\partial_{\mu} \Phi)^{\dag}(\partial_{\mu} \Phi ))^2~\Gamma^{2,1}[\Phi^{\dag}\Phi]+\cdots,
\eea
where $V(\Phi^{\dag}\Phi)$ is the effective potential and $\Gamma^{n,i}[\Phi^{\dag}\Phi]$ denotes the coefficient of the $i^{\rm th}$ term of $2n^{\rm th}$ derivative terms. 
The mass dimension of $\Gamma^{n,i}[\Phi^{\dag}\Phi]$ is equal to $4-4n$.
By setting $\Phi \to(0, (v+h)/\sqrt{2})$, the $q^2$ term of the physical Higgs ($h$) three-point function stems from
\bea
(\partial_{\mu} \Phi)^{\dag}(\partial_{\mu} \Phi )~\Gamma^{1,1}[\Phi^{\dag}\Phi]
&\ni&  \left. \frac{\delta \Gamma^{1,1}[(v+h)^2/2]}{\delta h} \right|_{h\to 0}(\partial_{\mu} h)^2h, \label{eq:31}\\
\Phi^{\dag}(\partial^2 \Phi )\Gamma^{1,2}[\Phi^{\dag}\Phi]
&\ni&  \left. \frac{\delta \Gamma^{1,2}[(v+h)^2/2]}{\delta h} \right|_{h\to 0}(\partial^2 h)h^2. \label{eq:32}
\eea
Hence, the $q^2$ term of three point function is defined by $\Gamma^{1,i}$. On the other hand, $\left. \Gamma^{1,i}[\Phi^{\dag}\Phi]\right|_{\Phi \to(0, (v+h)/\sqrt{2})} $ determine the Higgs wave function renormalization $Z$,
\be
{\mathcal L}\ni \frac{1}{2}Z (\partial_{\mu}h)(\partial_{\mu}h).
\ee
This means that the $q^2$ term of the three-point function is controlled by the wave function renormalization.\par
In a general CSI-model, the form of $\Gamma^{1,i}$ is fixed at LO by the following argument.
Since the bare Lagrangian has classical scale invariance, the effective action has only the classical field $\Phi$ and renormalization scale $\mu^2$ as dimensionful parameters, and since $\mu^2$ always appears in the logarithm, the form of $\Gamma^{1,i}$ is given by
\be
\Gamma^{1,i}[\Phi^{\dag}\Phi]=a^{i}+b^{i} \ln (\Phi^{\dag}\Phi/\mu^2), \label{eq:fnform}
\ee
where $a^i$ and $b^i$ represent dimensionless constants determined by coupling constants.
$a^i$ does not contain $\Phi$, only the logarithmic term contributes to eqs.(\ref{eq:31},\ref{eq:32}).
Furthermore, $\ln(\mu^2)$ is associated with the $1/\epsilon$ part of the wave function renormalization.\par
The above argument does not apply to models with dimensionful parameters such as $m^2$, since polynominal of $\Phi^{\dag}\Phi/m^2$ contributes to eq.(\ref{eq:fnform}). 
%This is finite part of wave function renormalization.
\par
In this paper, we consider a CSI model with SM gauge singlet scalar bosons.
Since the VEV of the singlet bosons is equal to zero, the previous argument applies and the contribution of the singlet-loops is expressed by eq.(\ref{eq:fnform}). 
Noting that the wave function renormalization by the singlet-loop is finite $(a^i<\infty,~b^i=0)$
%%%%%
\footnote{
$a^i$ is non-zero by the on-shell Higgs mass condition.
}
%%%%%
, the singlet bosons do not contribute to the $q^2$ term of the three-point function\footnote{
It is also possible to show this feature by explicit calculation even for a more general singlet sector with non-universal portal coupling or mixing among the singlet bosons. The mass eigenstates of the singlet scalars are also interaction eigenstates at LO since both eigenstates make $\lambda_{\rm HS}%H^{\dag}H \vec{S}\cdot\vec{S}
$ diagonalized at LO. We can see that the $q^2$ term cancels in each scalar component by eq.(\ref{eq:cancel}).
}.
\par
On the other hand, the fermion and vector bosons, {\it e.g.} the top, $W$ and $Z$, contribute to the divergent part of the Higgs wave function renormalization since the mass dimension of a fermion propagator is equal to $-1$ and $VVh$ has  derivative couplings.
As a result, fermions and vector bosons generally contribute to the $q^2$ term of the Higgs three-point function.
%$S$ matrix does not depended on gauge and wave function renormalization of internal particle, fermion and vector boson contribution for $q^2$ term of three point function will be canceled at $S$ matrix level.

\end{document}